\documentclass[preprint,superscriptaddress,aps,amssym]{revtex4}

\usepackage{latexsym}
\usepackage{textcomp}
\usepackage{graphicx,wrapfig}
\usepackage{epsfig}
\usepackage{dcolumn}
\usepackage{amsmath}

\usepackage{longtable}
\usepackage[usenames]{color}
\usepackage{amsfonts}
\usepackage{amssymb}

\usepackage{natbib}
\setcitestyle{numbers,sort&compress}

\usepackage[normalem]{ulem}
\usepackage{hyperref}
\usepackage{epstopdf}
\usepackage{float}
\usepackage{natbib}
\usepackage{multirow}
\usepackage{verbatim}

\input colordvi

\def\N2#1{\textcolor{blue}{N: {\em #1}}}
\def\N#1{\textcolor{red}{#1}}

\begin{document}

\title{Selective capture of ${\rm CO_2}$ over ${\rm N_2}$ and ${\rm CH_4}$: B clusters and their size effects}
\author{Alexandra B. Santos-Putungan}
\email[Author to whom correspondence should be addressed. Electronic address:] {absantos1@up.edu.ph}
\affiliation{Institute of Mathematical Sciences and Physics, University of the Philippines Los Ba\~{n}os, College, Los Ba\~{n}os, Laguna, Philippines}
\affiliation{Materials Science and Engineering Program, University of the Philippines Diliman, Quezon City, Philippines}

\author{Nata\v{s}a Stoji\'c}
\author{Nadia Binggeli}
\affiliation{The Abdus Salam International Centre for Theoretical Physics, Trieste, Italy}

\author{Francis N.~C.~Paraan}
\affiliation{National Institute of Physics, University of the Philippines Diliman, Quezon City, Philippines}

\begin{abstract}
  Using  density-functional theory (DFT), we investigate the selectivity of adsorption of CO$_2$ over ${\rm N_2}$ and  ${\rm CH_4}$ on planar-type B clusters, based on our previous finding of strong chemisorption of ${\rm CO_2}$ on the B$_{10-13}$ planar and quasiplanar clusters. We consider the prototype B$_8$ and B$_{12}$ planar-type clusters and perform a comparative study of the adsorption of the three  molecules on these clusters. We find that, at room temperature, ${\rm CO_2}$ can be separated from ${\rm N_2}$ by selective binding to the B$_{12}$ cluster and not to the B$_8$ cluster. Selective adsorption of ${\rm CO_2}$ over CH$_4$ at room temperature is possible for both clusters. Based on our DFT-adsorption data (including also  a semi-infinite Boron sheet) and the available literature-adsorption value for ${\rm N_2}$ on the planar-type B$_{36}$ cluster, we discuss the selectivity trend of ${\rm CO_2}$ adsorption over ${\rm N_2}$ and CH$_4$ with planar-cluster size, showing that it extends over sizes including B$_{10-13}$ clusters and significantly larger.\\

\end{abstract}

\startpage{1}
\endpage{ }
\maketitle

\section{Introduction}\label{intro}

Carbon-dioxide removal is essential for air purification and for minimizing the effects of global warming. Therefore, methods for elimination or reduction of ${\rm CO_2}$ from industrial processes are extremely important. Various kinds of gas mixtures are targeted for ${\rm CO_2}$ capture and separation\cite{LeyCarMar14,MonBalVar12}, e.g., postcombustion (predominantly ${\rm CO_2}$/${\rm N_2}$ separation) and natural gas sweetening (${\rm CO_2}$/${\rm CH_4}$ separation).

One approach to ${\rm CO_2}$ capture is to utilize selective adsorption properties of certain materials. In this respect, recent studies have investigated the selective adsorption of  ${\rm CO_2}$ by various materials including metal-organic frameworks \cite{Alezi2015, Saha2010, Ding2016}, nanomaterials \cite{Kunkel2016}, nanostructures \cite{Dai2009,Oh2014} and clusters \cite{SunWanLi14,Wang2015,Dong2015,Gao2015}. An important aspect is to identify materials which can selectively capture ${\rm CO_2}$ at room temperature. This requires, in particular, a strong adsorption energy, whose magnitude should typically overcome that of the standard free energy of gaseous ${\rm CO_2}$ at 300~K (0.67~eV) \cite{Choi2011}. A number of recent theoretical studies have revealed potentially interesting chemisorption properties of ${\rm CO_2}$  on Boron-based clusters and related model solid surfaces \cite{SunWanLi14,Wang2015,Dong2015,Gao2015,note_ourpaper,PhamPham2019,Sun2014PCCP}.  Small Boron clusters, containing 13 or less atoms, have been shown both experimentally and theoretically to be planar or quasiplanar \cite{Zhai2003,Oger2007,noteTai,Boustani1997}, while a number of medium size clusters (up to $\sim 40$ atoms) are also expected to occur in planar-type structures on the basis of available experimental and/or theoretical studies \cite{PiaHiLi14,Li2014JACS,Romanescu2012,Sergeeva2014,Pham2019}. Most of these clusters were produced using laser vaporization of a disk target \cite{Zhai2003,PiaHiLi14,Li2014JACS,Romanescu2012,Sergeeva2014}, and their corresponding photoemission  \cite{Zhai2003,PiaHiLi14,Li2014JACS,Sergeeva2014} and infrared absorption spectra \cite{Romanescu2012} were obtained.
Very recently, based on {\it ab initio} calculations, we have shown strong chemisorption of ${\rm CO_2}$ on the Boron B$_{10}$-B$_{13}$ planar or quasiplanar clusters, with magnitude of the chemisorption energy  surpassing 1~eV \cite{note_ourpaper}. We also found that the planarity of these clusters is the key for their particularly strong  ${\rm CO_2}$ chemisorption, and  that even in the limit of a semi-infinite B 2D sheet structure ${\rm CO_2}$ still chemically binds at 0 K, but with reduced strength \cite{note_ourpaper}.

So far, however, the selectivity of ${\rm CO_2}$ against other gas-molecule adsorption on the planar-type B clusters has not been investigated to the best of our knowledge. There have been some recent studies though, using density-functional theory (DFT), of gas-molecule  adsorption, including ${\rm CO_2}$ adsorption, on 3D fullerene-type  B$_{40}$ and  B$_{80}$ structures \cite{Dong2015,SunWanLi14}. B$_{40}$ is the only experimentally known Boron fullerene-type structure, while the B$_{80}$ fullerene had been proposed theoretically, but was not confirmed experimentally \cite{Dong2015}.  The calculated magnitude of the ${\rm CO_2}$ adsorption energy on B$_{40}$ (B$_{80}$) was  0.64 eV (0.84~eV), while for ${\rm N_2}$ and ${\rm CH_4}$ the magnitude was significantly smaller, being at most 0.16 and 0.26~eV, respectively. A DFT study has also addressed the adsorption of diatomic molecules, including N$_2$, on the B$_{36}$ quasiplanar cluster, and found that ${\rm N_2}$ does not bind at all to the cluster (has a positive adsorption energy) \cite{ValFarTab15}.

In this work we address the selectivity of ${\rm CO_2}$ adsorption over ${\rm N_2}$ and ${\rm CH_4}$  on planar B clusters. Using {\it ab initio} DFT calculations, we present first a comparative study of the adsorption of the three gas molecules on the B$_{12}$ and B$_8$ planar-type clusters. Our results indicate that, while B$_{12}$ displays perfect selectivity by binding ${\rm CO_2}$, but not N$_2$ and CH$_4$ at room temperature, the B$_8$ cluster cannot be used to separate ${\rm CO_2}$ from ${\rm N_2}$, as it binds both. B$_8$ does not bind CH$_4$ though (at room temperature), and separation of ${\rm CO_2}$ from ${\rm CH_4}$ by adsorption still holds also for B$_8$. Based on our present results and those for  ${\rm CO_2}$ adsorption on the semi-infinite B sheet and planar clusters of other sizes, and considering also the previous result for ${\rm N_2}$ on B$_{36}$, we then discuss the size dependence of the selectivity of the ${\rm CO_2}$ capture on the planar-type B clusters.

\section{Methodology}\label{method}

Density functional theory (DFT) calculations were performed using the Quantum ESPRESSO package \cite{Giannozzi2009}. We carried out spin-unrestricted computations within the generalized gradient approximation (GGA), using the spin-polarized  Perdew-Burke-Ernzerhof (PBE) functional \cite{Perdew1996}. Scalar relativistic Vanderbilt ultrasoft pseudopotentials were employed. In some specific cases of weak or no adsorption with GGA, we also carried out calculations with dispersion-corrected DFT (DFT-D)  \cite{Grimme2006} to check the effect of van der Waals correction.

We considered Gamma-point Brillouin zone sampling with a Gaussian level smearing of 0.001~Ry. Atomic structures were optimized until the forces on each atom were below 10$^{-4}$~Ry/a.u. A large cubic cell of 25~\AA~$\times$~25~\AA~$\times$~25~\AA~was used to avoid interactions between periodic images of the cluster. The kinetic energy cut-off for the plane-wave expansion of the electronic orbitals and density were 60 and 240~Ry, respectively. Convergence tests indicated that the adsorption energy was converged within 1 meV with these parameter values.

The equilibrium structures used for the ground-state isomers of the Boron clusters were based on the study of Tai et al \cite{Tai2010} and reproduced by us. The electronic ground states we obtain for the B$_{12}$ and B$_{7}$ clusters, based on the analysis of the molecular orbitals, correspond to the $^1$A$_1$(C$_{3v}$)   and  $^3$A'$_2$(D$_{7h}$) ground states of previous studies \cite{Tai2010,Zubarev2007}.
 The adsorption energy, $E_\textrm{ads}$, of the molecule (CO$_2$, N$_2$ and CH$_4$) on the Boron cluster was computed as:
\begin{equation} \label{eq:E_ads}
E_\textrm{ads}=E_{\textrm{B}_{n}-\textrm{gas}}-{E}_{\textrm{B}_{n}}-{E}_\textrm{gas},
\end{equation}
where $E_{\textrm{B}_n-\textrm{gas}}$ is the total energy of the atomically relaxed system consisting of the B$_{n}$ cluster and adsorbed gas molecule, $E_{\textrm{B}_n}$ is the total energy of the isolated (relaxed) B$_{n}$ cluster (for the ground-state isomer, unless otherwise specified), and $E_{\textrm{gas}}$ is the total energy of the gas molecule (CO$_2$, N$_2$ and CH$_4$ in gas phase).

Different cluster adsorption sites (see also Appendix) and different molecule orientations were considered as starting configurations for the adsorption of each molecule. Furthermore, in the cases in which there was no strong binding at $0$~K of the molecule to the ground-state isomer of the cluster (E$_\textrm{ads} > - 0.6$~eV), we also considered as cluster starting structure for the adsorption relaxations the second lowest-energy isomer. For the B$_{12}$ cluster, which has as ground state a quasi-planar (buckled) structure, we considered also the planar (D$_{3h}$) isomer, which is 0.19 eV higher in energy. For B$_8$, whose ground-state isomer is the triplet (spin moment $m = 2 \mu_B$) heptagon structure, we also examined the singlet  ($m=0$) wheel-type isomer \cite{LiZhaoXuLi2005}, which lies 0.48 eV higher in energy.  In all cases, however, the resulting lowest-energy B$_n$-molecule configuration found was either higher in energy than that found starting from the ground-state isomer or resulting essentially in the same lowest-energy B$_n$-molecule configuration (see Appendix).  In the following, therefore, we will focus on the results obtained starting from the B$_n$ ground-state isomer.

\section{Results and Discussion}\label{RandD}

\subsection{Adsorption on the B$_{12}$ cluster}

\begin{figure}[!h]
\centering\includegraphics[width=0.7\linewidth]{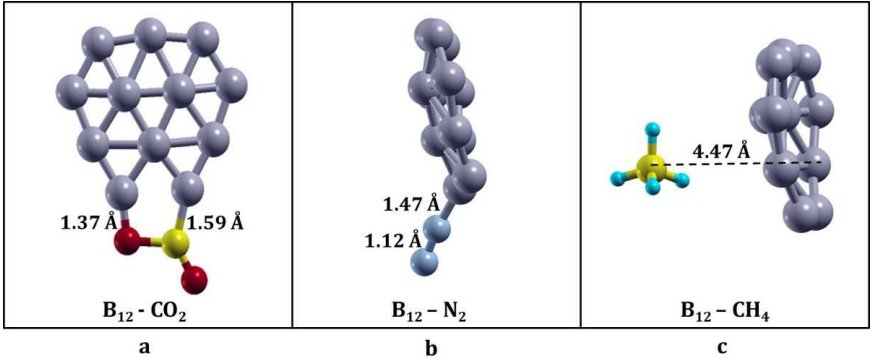}
\caption{Optimized structures of CO$_2$, N$_2$ and CH$_4$ molecules adsorbed on B$_{12}$ cluster shown for the configurations  characterized by the strongest adsorption found for each gas molecule.  B, C, O, N and H atoms are shown in grey, yellow, red, navy blue and light blue, respectively.}
\label{fig:B12-gas}
\end{figure}

Figure~\ref{fig:B12-gas} shows the optimized configurations for the gas molecules adsorbed on the B$_{12}$  cluster, characterized by the strongest adsorption energies we obtain from all the relaxation calculations. These final configurations correspond to molecules initially placed at a distance of 2~\AA\ from the cluster, with an in-plane orientation for CO$_2$ and N$_2$, while CH$_4$ was out-of-plane. As reported previously \cite{note_ourpaper}, the CO$_2$ molecule is strongly chemisorbed on the cluster (E$_\textrm{ads}$ = $-1.6$~eV). On the other hand, we find that the N$_2$ molecule (at T = 0) binds relatively weakly to the B$_{12}$ cluster (compared to the CO$_2$ molecule), having an adsorption energy of only $-0.13$~eV, while no binding (a positive adsorption energy) is found for CH$_4$ on the cluster within GGA. Table \ref{tab:B12table} summarizes the adsorption energy corresponding to the strongest binding we find in GGA for each of the gas molecules considered, together with the main bond lengths and angles associated with the different molecules.

\begin{table}[!h]
\centering
\caption{Calculated adsorption energies (E$_\textrm{ads}$) within GGA-PBE for the gas molecules adsorbed on B$_{12}$, together with the bond lengths and angles for the configurations corresponding to the strongest adsorption found (for ${\rm CO_2}$ the angle corresponds to the bending of the molecule, while for the ${\rm N_2}$ it takes into account also the bond with B). The \lq\lq standard free energy\rq\rq of gaseous CO$_2$ ($-S_{\rm gas}^\circ T$) at 300~K (see text) is also given for comparison (third column).}
\begin{tabular}{c  c  c  c  c}
\hline
\textbf{Gas molecule} & \textbf{E$_\textrm{ads}$ (eV)} & \textbf{\lq\lq Standard free energy\rq\rq} & \textbf{Interatomic distance from} & \textbf{Angles } \\
 &  & \textbf{($-S_{\rm gas}^\circ T)$ (eV)} & \textbf{nearest B atom (\AA)  } &   \\
\hline
CO$_2$ & $-1.60$$^{\text{\cite{note_ourpaper}}}$ & $-0.67$$^{\text{\cite{Choi2011}}}$ & O\textendash B $1.37$ & $\angle$OCO $122^{\circ}$ \\
    &   &  &  C\textendash B $1.59$ &  \\
\hline
N$_2$ & $-0.13$ & $-0.60$$^{\text{\cite{entropy_N2}}}$ & N\textendash B $1.47$ & $\angle$NNB $171^{\circ}$ \\
\hline
CH$_4$ & $+0.01$ & $-0.58$$^{\text{\cite{entropy_N2}}}$ &  C\textendash B $4.47$ & \textemdash \\
\hline
\end{tabular}
\label{tab:B12table}
\end{table}

For the chemisorbed B$_{12}$\textendash CO$_2$ system, the adsorbed CO$_2$ molecule lies in the plane of the cluster (Figure~\ref{fig:B12-gas}a). This favors strong interaction between the in-plane lowest unoccupied molecular orbital (LUMO) of the bent CO$_2$ and the in-plane high-energy occupied molecular orbitals of the B cluster \cite{note_ourpaper}. The O\textendash B and C\textendash B distances are found to be 1.4 and 1.6~\AA, respectively. The (initially linear)  CO$_2$ molecule bends at an angle of 122$^{\circ}$ as it chemisorbs on the B$_{12}$ cluster, which corresponds to the equilibrium geometry predicted for the negatively charged CO$_2$ molecule \cite{GutBarCom98}. We note that the strong chemisorption of   CO$_2$ on  B$_{12}$ is also robust within the GGA+U approach, which increases the  CO$_2$ HOMO-LUMO band gap \cite{note_ourpaper}.

For the B$_{12}$\textendash N$_2$ system, the adsorbed N$_2$ lies outside the cluster plane (Figure~\ref{fig:B12-gas}.b). Aside from its weak adsorption energy, we can observe a N-B distance of 1.47~\AA. The N$_2$ bond length barely changes upon adsorption (from 1.10~\AA\ to 1.12~\AA\ for isolated to adsorbed N$_2$). Including long-range dispersion correction for  B$_{12}$\textendash N$_2$ only slightly increases the adsorption strength, changing  E$_\textrm{ads}$ from -0.13 eV (DFT) to -0.24 eV (DFT-D). It is interesting to compare our results for the N$_2$ adsorption on the ground-state planar-type isomer of B$_{12}$ to the results of a DFT-D study of ${\rm N_2}$ adsorption on the 3D cage-type icosahedral B$_{12}$ system \cite{SunWanLi13} (which is not the cluster ground state, but a building element of the bulk $\alpha-B_{12}$ and $\gamma-B_{28}$ phases of solid Boron and its surfaces). This  reveals that ${\rm N_2}$ adsorbs more strongly, with E$_\textrm{ads}$ = $-0.38$~eV \cite{SunWanLi13}, on the 3D cage-like icosahedral  $B_{12}$. For the larger B$_{40}$ and B$_{80}$ cage-like structures, DFT-D studies indicate that, with increasing cluster size, the $N_2$ adsorption-energy amplitude is reduced respectively to 0.16 eV and 0.14 eV \cite{Dong2015,SunWanLi14}, becoming comparable to the amplitude of our DFT-D value on the smaller planar-type B$_{12}$ cluster (0.24~eV). For the larger B$_{36}$ planar-type cluster, a recent DFT study with B3LYP functional also indicates a decreased adsorption of N$_2$ with the increased size (in that case, a slightly positive value) \cite{ValFarTab15}.

For CH$_4$, the calculated adsorption energy on the planar B$_{12}$ cluster is slightly positive (nearly zero). We checked the effect of the van der Waals correction using DFT-D, which is to  induce a very weak adsorption of $-0.13$~eV. In the DFT-D study for CH$_4$ on the B$_{40}$ (B$_{80}$) cage-like structure, the adsorption energy was also negative, but somewhat stronger in magnitude, with E$_\textrm{ads}$ of -0.26 eV (-0.14 eV) \cite{Dong2015,SunWanLi14}.

Considering the \lq\lq standard free energy\rq\rq at 300 K (consisting of the entropy term $-S_{\rm gas}^\circ T$) \cite{Choi2011} of the gaseous CO$_2$ (-0.67 eV)  \cite{Choi2011} and N$_2$ (-0.60 eV) \cite{entropy_N2}, whose  magnitude (absolute value) should be an upper bound for the entropy contribution to the adsorption free energy $-\Delta S\cdot T$ \cite{Choi2011,SunLi2013} $(0 < -\Delta S \cdot T < +S_{\rm gas}^{\circ}T)$, it is evident from Table \ref{tab:B12table} that the CO$_2$ molecule will remain attached to the cluster even at room temperature. In the case of the gas molecules which bind at $0~K$ (E$_\textrm{ads} < 0$) to planar-type B clusters such as B$_{36}$, the molecule gaseous free-energy contribution ($S_{\rm gas}^\circ \cdot T$) dominates the entropy contribution to the adsorption free-energy ($-\Delta S \cdot T$) \cite{ValFarTab15}. Hence, for N$_2$, as the DFT (DFT-D) binding energy of 0.13 eV (0.24~eV) to B$_{12}$ is considerably smaller compared to the magnitude of the standard free energy of gaseous N$_2$, the molecule will not stick to the cluster at room temperature. Similarly, CH$_4$ will not stick either at room temperature. Hence, our results indicate selective capture of CO$_2$ against N$_2$ and CH$_4$ by the B$_{12}$ planar-type cluster at room temperature. Clearly at very low temperature ($T \approx 0$~K), N$_2$ would bind to the cluster (and also CH$_4$, with van der Waals interaction).

\subsection{Adsorption on the B$_{8}$ cluster}

\begin{figure}[!h]
\centering\includegraphics[width=0.7\linewidth]{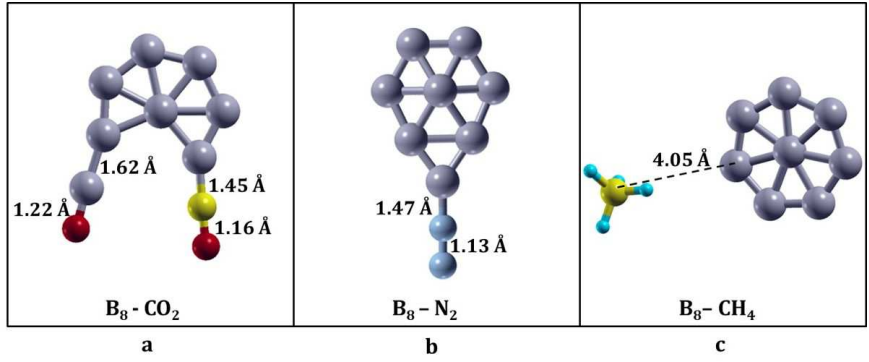}
\caption{Obtained optimized structures for different gas molecules (CO$_2$, N$_2$ and CH$_4$) adsorbed on B$_8$ cluster resulting in the strongest adsorption for each molecule. For B$_8$\textendash CO$_2$ (a), a dissociated chemisorption can be observed. Chemisorption of N$_2$ molecule is found (b), while a very weak physisorption is observed for CH$_4$ molecule on B$_8$ cluster (c). The atom colors are like in Fig~\ref{fig:B12-gas}.}
\label{fig:B8-gas}
\end{figure}

In Figure~\ref{fig:B8-gas}, we show the lowest-energy configurations obtained for the three gas molecules adsorbed on the planar B$_8$ cluster with the same initial orientation and distance from the cluster as in the case of the adsorption on B$_{12}$. The corresponding adsorption energies and bond lengths/angles are reported in Table~\ref{tab:B8table}. It can be observed that the adsorption energy is systematically lower (stronger binding) for the three molecules on B$_8$, compared to B$_{12}$.

For CO$_2$, we find strong chemisorption with dissociation of CO$_2$, as shown in Fig.~\ref{fig:B8-gas}a. We note that this chemisorption is stronger  (E$_\textrm{ads}$ = $-2.40$~eV) than found earlier for the doped CoB$_8^{-}$ cluster \cite{Wang2015}, on which the molecule was not dissociating. When chemisorbed on B$_8$, the CO$_2$ molecule dissociates spontaneously (without an energy barrier) in our relaxation calculations. The resulting chemisorbed system has a planar geometry. One can notice (in Fig.~\ref{fig:B8-gas}a) that some of the B\textendash B cluster bonds were broken to accommodate new bonds with the O atom and the CO molecule  (of the dissociated CO$_2$), as we also observed previously from DFT calculations for two of the larger planar B clusters, B$_{11}$ and B$_{13}$ \cite{note_ourpaper}. A dissociative CO$_2$ adsorption was also found in a recent DFT study for CO$_2$ adsorbed on B$_7^{-}$ and V$_2$B$_7$ clusters \cite{PhamPham2019}, with also largely enhanced amplitude of the adsorption energy, especially for the negatively charged cluster B$_7^{-}$ \cite{PhamPham2019}.

For N$_2$, we obtain also rather strong adsorption to B$_8$ (E$_\textrm{ads}$ = $-0.76$~eV); it is more than 5 times stronger than in the case of B$_{12}$. As shown in Fig.~\ref{fig:B8-gas}b, adsorption of the molecule also causes atomic rearrangement of the cluster, though it retains a compact geometry (the Boron 8-atom centered-wheel structure rearranges forming a reduced 7-atom centered-wheel quasi-planar structure, by removing a B atom from the original 7-atom circle to bind with N$_2$). The rearranged wheel-type cluster shows a 2-fold symmetry. Similar to the N$_2$-B$_{12}$ case, the bond length of the N$_2$ molecule adsorbed on B$_8$ only slightly increases (from 1.10 to 1.13~\AA), despite a considerable increase in binding energy. Also the bending angle $\angle$NNB and the B-N bond length are virtually the same as for the adsorption on B$_{12}$.

For CH$_4$ we find a very weak binding (physisorption) on B$_8$. The lowest-energy configuration we obtain is shown in Fig.~\ref{fig:B8-gas}c, and is characterized by a C\textendash B distance of 4.05~\AA\ and E$_\textrm{ads} = -0.01$~eV. Including van der Waals correction within DFT-D only slightly changes  E$_\textrm{ads}$ to $-0.10$~eV.

\begin{table}[!h]
\centering
\caption{Calculated lowest adsorption energies (E$_\textrm{ads}$) within GGA-PBE obtained for the three gas molecules on the B$_{8}$ cluster, together with the corresponding bond lengths and angle.
}
\begin{tabular}{c  c  c  c}
\hline
\textbf{Gas molecule} & \textbf{E$_\textrm{ads}$ (eV)} & \textbf{Interatomic distance from} & \textbf{Angles } \\
 &  & \textbf{nearest B atom (\AA)} &   \\
\hline
CO$_2$ & $-2.40$ & O\textendash B $1.22$ &  $\angle$OCO $122^{\circ}$\\
    &   &  C\textendash B $1.45$ &  \\
\hline
N$_2$ & $-0.76$ & N\textendash B $1.47$ & $\angle$NNB $171^{\circ}$ \\
\hline
CH$_4$ & $-0.01$ & C\textendash B $4.05$ & \textemdash \\
\hline
\end{tabular}
\label{tab:B8table}
\end{table}

For the B$_8$ planar cluster thus the magnitude of the adsorption energy of the N$_2$ molecule (0.76 eV) is well above the magnitude of the standard free energy of the gaseous N$_2$ (0.60~eV). Hence, unlike B$_{12}$, for B$_8$ both CO$_2$ and N$_2$ are expected to stick to the cluster also at room temperature.

We note that the much higher reactivity of B$_8$, compared to B$_{12}$, which even leads to dissociated adsorption of CO$_2$ on B$_8$, can be understood comparing the electronic molecular-orbital structure of the two clusters  \cite{Zubarev2007}. Unlike B$_{12}$, which  has a large HOMO-LUMO gap (2.3 eV in our calculations), the D$_{7h}$ - B$_8$ is characterized by orbital degeneracy at the Fermi enery (an open-shell/``metallic'' character) with half-filled two-fold degenerate 1e$_1$'' orbitals \cite{Zubarev2007} in non-spin polarized DFT calculations. In the spin-polarized calculations, the degeneracy is lifted (with a small gap) by the formation of a magnetic ground state of moment $2 \mu_B$ \textendash  corresponding to the triplet D$_{7h}$ - $ ^3$A'$_2$ ground state \cite{Tai2010,Zubarev2007}. The HOMO-LUMO gap, however, remains small (1.0 eV in our calculations) compared to that of B$_{12}$.  We attribute thus the much higher  reactivity of B$_8$  (leading to dissociated adsorption of CO$_2$) predominantly  to its near orbital degeneracy. In fact, it should be noted that the other B clusters for which we found earlier \cite{note_ourpaper} dissociated adsorption of CO$_2$, i.e., B$_{11}$ and B$_{13}$ (unlike B$_{10}$ and B$_{12}$) are clusters which all display a zero HOMO-LUMO gap or ``metallic'' character (having an odd number of electrons) in calculations neglecting spin polarization  \textendash  and are characterized by small gaps (much smaller than those of B$_{10}$ and B$_{12}$) in the spin-polarized calculations (see Supplementary Material of Ref. \onlinecite{note_ourpaper}).  Apart from this factor, however, we also observe an overall smoother general tendency of decreasing binding energy $|E_\textrm{ads}|$ of the adsorbed molecules with increasing size of the B clusters, as will be illustrated in the next section.

\subsection{Size dependency of the selective adsorption}

As we have seen in the previous sections, selective binding of CO$_2$ over N$_2$ and CH$_4$ can be expected for the B$_{12}$ cluster at room temperature, but not for B$_8$. B$_8$ is predicted to bind, instead, both N$_2$ and CO$_2$ at room temperature (against CH$_4$). We also observed a decrease in the adsorption binding strength, for all three molecules, from B$_8$ to B$_{12}$.

\begin{figure}[!h]
\centering\includegraphics[width=1.0\linewidth]{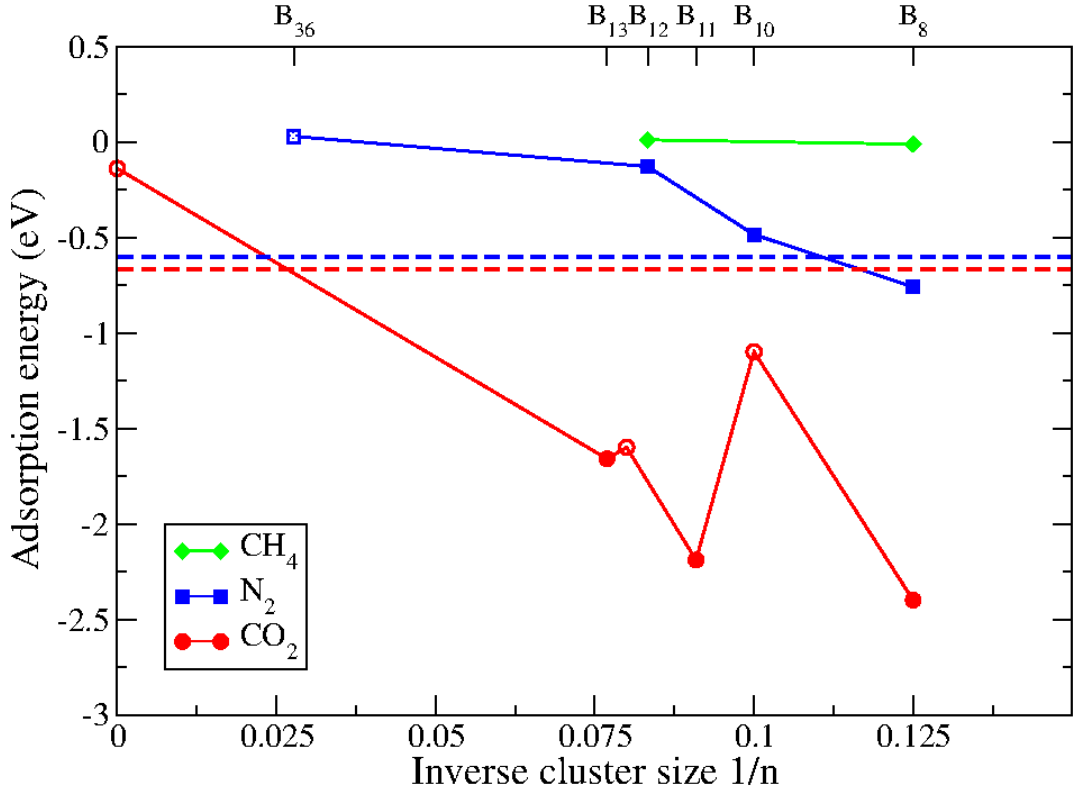}
\caption{The lowest adsorption energies of CO$_2$, N$_2$ and CH$_4$ on the B$_n$ planar-type clusters obtained within GGA-PBE (except for N$_2$ on B$_{36}$  calculated with B3LYP \cite{ValFarTab15}). The values for CO$_2$ on B$_{10-13}$ and on the 2D Boron sheet ($n\rightarrow\infty$) are taken from our previous calculations \cite{note_ourpaper}. The open symbols for the CO$_2$ adsorption stand for the single-molecule adsorption (and full symbols for dissociated molecule) and for the N$_2$ case for the literature value \cite{ValFarTab15}. The dashed lines represent the standard free energy of the gaseous CO$_2$ (red) and N$_2$ (blue) \cite{entropy_N2} at 300~K.}
\label{fig:ads}
\end{figure}

In Figure~\ref{fig:ads}, we show the  available lowest-energy data from GGA/DFT calculations (B3LYP calculations for N$_2$ on B$_{36}$) obtained for each of the three molecules on the B$_n$ planar-type clusters, as a function of  inverse cluster size 1/n. Apart from the results for the three molecules on B$_8$ and B$_{12}$, we  also included  the lowest $\rm E_{ads}$ values for CO$_2$ on the B$_{10}$, B$_{11}$, B$_{13}$ and on the semi-infinite B sheet ($n\rightarrow\infty$)  from our previous calculations  \cite{note_ourpaper} and  the value of Ref.~\onlinecite{ValFarTab15} for the adsorption of N$_2$ on the planar-type B$_{36}$ cluster. We note that for the CO$_2$ adsorption energy we used two different symbols to distinguish the cases in which the calculated final relaxed adsorbates correspond to the single (bent) CO$_{2}$  molecule or to the dissociated CO$_2$ (which leads to lower $\rm E_{ads}$). We also indicated, in Figure~\ref{fig:ads}, the values of the standard free energy at $T=300$~K of the gaseous CO$_2$ (-0.67~eV) and N$_2$ (-0.60~eV) (the value for gaseous CH$_4$ is -0.58~eV, \cite{entropy_N2}).

In our previous study \cite{note_ourpaper}, we found that the adsorption strength weakens for ${\rm CO_2}$ adsorption on a B 2D sheet, with respect to adsorption on B$_{10-13}$ clusters, remaining, however, significant and calculated to be -0.14~eV. Our overall  results on CO$_2$ adsorption on the B$_n$ clusters, in Figure~\ref{fig:ads}, display a general tendency of decreasing ${\rm CO_2}$ adsorption strength with increasing size $n$ of the planar B clusters.
A similar trend of decreasing adsorption strength on the B planar clusters with increasing cluster size is found for the ${\rm N_2}$ molecule, for which we have also evaluated the adsorption energy within GGA-PBE for the planar B$_{10}$ cluster (E$_\textrm{ads}$ = $-0.49$~eV). However, the magnitude of the overall change and oscillations with cluster size are much larger for ${\rm CO_2}$. From the literature we know that for the even larger planar cluster, B$_{36}$, the adsorption energy (within DFT B3LYP) is positive (0.03 eV) \cite{ValFarTab15}, indicating that ${\rm N_2}$ changes from binding at room temperature on B$_8$ to not binding at any temperature for B$_{36}$. In the case of the CH$_4$ molecule, the adsorption strength also decreases with increasing cluster size, but the variation is very weak  (a few meV from B$_8$ to B$_{12}$).

Considering the standard free-energy values at T = 300 K of the gaseous CO$_2$ and N$_2$ (-0.67 eV and -0.60 eV, respectively), and seeing how large (in Figure~\ref{fig:ads}) the $| E_\textrm{ads}|$ values of ${\rm CO_2}$ are on the planar B$_{10-13}$ clusters and how rapidly the ${\rm N_2}$ adsorption strength decreases with cluster size, we can expect  that all B$_{10-13}$ clusters should selectively bind CO$_2$ against N$_2$ at room temperature, and could be used thus to separate ${\rm CO_2}$ from ${\rm N_2}$. Moreover, taking into account our calculated ${\rm CO_2}$ adsorption strength on a B 2D sheet and knowing there is no ${\rm N_2}$ adsorption on  B$_{36}$ \cite{ValFarTab15}, we find it very likely (from Figure~\ref{fig:ads}) that the selective adsorption of ${\rm CO_2}$ over ${\rm N_2}$ at room conditions can be extended to significantly larger B planar-cluster sizes, possibly up to the limit ($n < 40$) beyond which the B clusters display a 3D-type structure.
Similarly, regarding the selectivity of ${\rm CO_2}$ over ${\rm CH_4}$, we expect adsorption at room (and low) temperature on all planar B clusters up to the same size limit to be selective, in view of the nearly vanishing E$_\textrm{ads}$ of ${\rm CH_4}$ on the planar B clusters.

\section{Conclusion}

We have investigated the selectivity of the adsorption of ${\rm CO_2}$ over ${\rm N_2}$ and ${\rm CH_4}$ on planar-type B clusters. We carried out a first-principles DFT comparative study of the adsorption of ${\rm CO_2}$, ${\rm N_2}$ and ${\rm CH_4}$ on the B$_8$ and B$_{12}$ planar-type clusters. Strong chemisorption (stable at room temperature) of ${\rm CO_2}$ is found on both clusters. At the same time, the ${\rm N_2}$ adsorption is strong (stable at room temperature) on B$_8$, while it is weak on B$_{12}$. We find that ${\rm CH_4}$ physisorbs at T = 0~K on B$_8$ and on B$_{12}$ (including van der Waals interaction), but does not bind at room temperature.  This implies that selective adsorption of ${\rm CO_2}$ over ${\rm N_2}$ at room temperature is possible for B$_{12}$ and not for B$_8$, while for both cluster sizes ${\rm CO_2}$ can be separated from ${\rm CH_4}$ at room temperature.  Moreover, based on the present  results, as well as our previous results on ${\rm CO_2}$ adsorption on B$_{10-13}$ and on a B 2D sheet, and the literature value of the ${\rm N_2}$ adsorption on B$_{36}$, we discussed the selectivity trend of the ${\rm CO_2}$ adsorption over ${\rm N_2}$ and CH$_4$ with B planar-cluster size, showing it extends over sizes including the B$_{10-13}$ clusters and significantly above.

\section{Acknowledgements}\label{acknowledgements}
This work was funded by the UP System Enhanced Creative Work and Research Grant ECWRG 2018-1-009. A.B.S.-P. is grateful to the Abdus Salam International Centre for Theoretical Physics (ICTP) and the OPEC Fund for International Development (OFID) for the OFID-ICTP postgraduate fellowship under the  ICTP/IAEA Sandwich Training Educational Programme, and to the Philippines Commission on Higher Education (CHEd) for the Faculty Development Program (FacDev)\textendash Phase II.

\section{Appendix: Adsorption sites and isomers}

In our study we have examined for each cluster (and for each considered isomer) different non-equivalent adsorption sites,  for all three molecules. The results corresponding to the lowest-energy adsorption configuration (and strongest binding) were reported in the paper and were obtained starting from the ground-state isomer.

\begin{figure}[H]
\centering\includegraphics[width=0.7\linewidth]{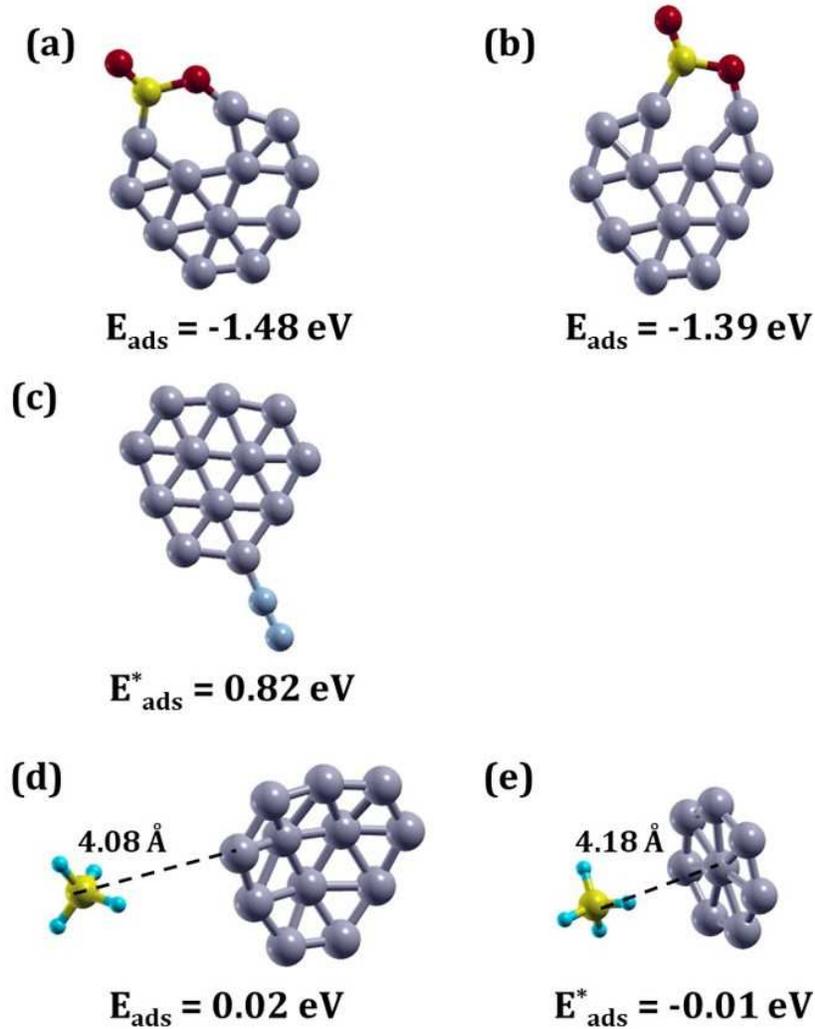}
\caption{Examples of other optimized structures for CO$_2$, N$_2$ and CH$_4$ adsorption on the Boron clusters.  Lowest-energy configurations obtained for CO$_2$ adsorbed at different inequivalent sites of the B$_{12}$  cluster (a,b).  Adsorption of N$_2$ on the flat B$^*_{12}$ isomer, when the cluster structure remains flat upon adsorption, i.e., when N$_2$ is initially placed in the plane of the cluster (c). Lowest-energy configuration obtained for CH$_4$ when adsorbed on the flat second-lowest-energy isomer (B$^*_{12}$) of the B$_{12}$ cluster (d) and on the second-lowest-energy isomer (B$^*_8$) of the B$_8$ cluster (e). In (d)  one nearly recovers the lowest-energy configuration obtained for CH$_4$ adsorbed on the ground-state B$_{12}$ isomer. In (e) the adsorption energy E$^*_{\textrm{ads}}$ (obtained using for E$_{\textrm{B}_n}$ the energy of the second lowest-energy isomer B$^*_8$) is the same as the value of  the adsorption energy E$_\textrm{ads}$ on the ground-state isomer, however, the energy of the B$^*_8$-CH$_4$ structure  is considerably larger (0.48 eV larger) than that of the B$_8$-CH$_4$ structure in Fig.~\ref{fig:B12-gas}(c). The atom colors are like in Fig.~\ref{fig:B12-gas}. }
\label{fig:OtherBn-gas}
\end{figure}

In Fig.~\ref{fig:OtherBn-gas} (a) and (b), we show, in  the case of CO$_2$ adsorption on the B$_{12}$ ground-state isomer, the results for the optimized configuration and adsorption energy obtained for some of the other inequivalent cluster adsorption sites giving strong binding (chemisorption). These adsorption configurations are rather similar to the lowest-energy case (Fig.~\ref{fig:B12-gas}(a)) and the binding energy ($|E_\textrm{ads}|$)  is only about 0.1 to 0.2 eV weaker. Based on a Lowdin charge analysis, which may be used only to get some indicative charging trends \cite{note_ourpaper}, we also find that the Lowdin charge gained by the CO$_2$ and lost by the cluster remains virtually the same (within $0.01$~$e$) for the configurations (a) and (b) in Fig.~\ref{fig:OtherBn-gas}, with respect to the lowest-energy configuration shown in Fig.~\ref{fig:B12-gas} (a), i.e., $\sim 0.30$~$e$ lost by the cluster and gained by the CO$_2$.

In the cases in which no strong binding (E$_\textrm{ads} > 0.6$ eV) was found for the molecule to the ground-state isomer of the cluster at 0 K (i.e., the cases of N$_2$ on B$_{12}$ and of CH$_4$ on B$_{12}$ and on B$_8$), we also considered, as starting structure of the cluster for adsorption, the second lowest-energy isomer (denoted B$^*_n$). For B$_{12}$, we considered the flat (D$_{3h}$) isomer, whose energy is 0.19 eV higher than the buckled isomer in our GGA calculations. This is of the same order of magnitude as the energy difference between the flat and buckled isomer of B$_{36}$ (a couple of tenths of eV) obtained in the B3LYP study \cite{ValFarTab15}. For B$_8$, we considered the singlet wheel-type structure, which lies 0.48 eV above the ground-state isomer in our GGA study, in good agreement with the B3LYP value of 0.51 eV for the  energy difference between these two B$_8$ isomers \cite{LiZhaoXuLi2005}.

For the adsorption of N$_2$, the optimized configurations we obtain starting from the flat B$^*_{12}$ isomer are higher in energy than those obtained starting from the buckled B$_{12}$. In particular, configurations in which the B$^*_{12}$ remains flat upon adsorption of N$_2$ (i.e., when N$_2$ is initially placed in the plane of the cluster) are highly unfavorable energetically, as illustrated in Fig.~\ref{fig:OtherBn-gas}(c).
Fig.~\ref{fig:OtherBn-gas}(d) shows the lowest-energy configuration we find for the CH$_4$ molecule adsorbed on the flat B$^*_{12}$ isomer. Starting from the flat B$^*_{12}$, we recover the same type of relaxed B$_{12}$-CH$_4$ configuration as found with the ground-state isomer (the flat isomer buckles) and the adsorption energy E$_\textrm{ads}$ (using for $E_{\textrm{B}_n}$ the energy of the ground-state isomer) is nearly the same (0.02 eV in GGA) as found starting with the ground-state isomer (0.01 eV).  In Fig.~\ref{fig:OtherBn-gas}(e), we show the lowest-energy configuration we find for CH$_4$ adsorbed on the second lowest-energy  B$^*_8$ isomer. The adsorption energy (E$^*_\textrm{ads}$) obtained from Eq.~\ref{eq:E_ads}  using for E$_{\textrm{B}_n}$  the energy of the second lowest-energy isomer B$^*_8$ is -0.01 eV, corresponding to the value of  E$_\textrm{ads}$ for the adsorption of the molecule on the ground-state isomer. The energy, however, of the B$^*_8$-CH$_4$  structure, in Fig.~\ref{fig:OtherBn-gas}(e), is considerably larger (by 0.48 eV) than that of the B$_8$-CH$_4$ system in Fig.~\ref{fig:B12-gas}(c). Hence, in all of these cases, the lowest-energy adsorbed molecule-cluster configuration obtained is either higher in energy than that found for the cluster ground-state isomer or resulting in nearly the same lowest-energy configuration.

It is interesting to observe, in Fig.~\ref{fig:B12-gas}(a) and in Fig.~\ref{fig:OtherBn-gas} (a) and (b), that the B$_{12}$ cluster flattens upon adsorption of CO$_2$ (which is not the case for N$_2$ and CH$_4$). This occurs in situations of strong adsorption, when one of the B-B bonds breaks, and is similar to the cases of strong adsorption of the NO and O$_2$ molecules on the B$_{36}$ cluster in the DFT/B3LYP study \cite{ValFarTab15}. We would like also to note, concerning the isomers and regarding the finite-temperature entropy term in the adsorption free energy ($- \Delta S \cdot T$), that the contribution due to the difference between the two isomers ($- \Delta S_\textrm{isomer} \cdot  T$) at room temperature (which is neglected in our study) is expected to be of the same order or less than that (the entropy-term difference $- \Delta S_\textrm{isomer} \cdot  T$) calculated for the two free B$_{36}$ isomers in Ref.~\onlinecite{ValFarTab15}, i.e., $\sim 0.2$~ eV. For our discussion on the sticking of the molecule at room temperature, this is of course very small (can be neglected) compared to the CO$_2$ chemisorption energy on the B$_{12}$ cluster and it is also small compared to the standard free energy of the gaseous molecule ($- S^0_{gas} \cdot  T$) at room temperature, included in our discussion. We would like also to emphasize that in the cases of molecules which are found to bind to  planar-type Boron clusters such as B$_{36}$ \cite{ValFarTab15}, the entropy term of the adsorption free energy ($- \Delta S \cdot T > 0$) at room temperature is also dominated by the molecule gaseous free-energy contribution ($+ S^0_{gas} \cdot  T$), and the difference between these two values (which are both positive, with the latter larger, i.e.,  $ + S^0_{gas} \cdot  T > - \Delta S \cdot T > 0$) is typically at most $0.2$~eV, based on previous studies \cite{ValFarTab15}. Such a difference does not modify the conclusions of our discussion on the sticking or non-sticking of the CO$_2$, N$_2$, and CH$_4$ molecules on B$_{12}$ and B$_8$.

\bibliography{Boron-Gas}

\bibliographystyle{unsrtnat}

\end{document}